\documentclass[aps,prb,twocolumn]{revtex4} 
\usepackage{amsmath} 
\usepackage{graphicx} 
\usepackage{color}

\begin{document} 
\title{Supercurrent-Induced Spin-Orbit Torques} 

\author{Kjetil M. D. Hals} 
\affiliation{Niels Bohr International Academy and the Center for Quantum Devices, Niels Bohr Institute, University of Copenhagen, 2100 Copenhagen, Denmark} 
\begin{abstract}
We theoretically investigate the supercurrent-induced magnetization dynamics of a two-dimensional lattice of ferromagnetically ordered spins placed on a conventional superconductor with broken spatial inversion symmetry and strong spin-orbit coupling. We develop a phenomenological description of the coupled dynamics of the superconducting condensate and the spin system, and demonstrate that supercurrents produce a reactive spin-orbit torque on the magnetization. By performing a microscopic self-consistent calculation, we show that the spin-orbit torque originates from a  spin-polarization of the Cooper pairs due to current-induced spin-triplet correlations.  
Interestingly, we find that there exists an intrinsic limitation for the maximum achievable spin-orbit torque, which is determined by the coupling strength between the condensate and the spin system.
In proximitized hole-doped semiconductors, the maximum achievable spin-orbit torque field is estimated to be on the order of $0.16$ mT, which is comparable to the critical field for current-induced magnetization switching in ferromagnetic semiconductors.   
\end{abstract}

\maketitle 

\section{Introduction} 
In metallic ferromagnets, spin-polarized currents can induce a spin-transfer-torque (STT) on the magnetization via direct transfer of spin-angular momentum from the itinerant charge carriers to the local magnetic moments.~\cite{Brataas:nm2012} 
This phenomenon has opened the door for current-driven manipulation of magnetization in spintronic devices. 
However, a limiting factor is the high current densities required to switch the magnetization and the associated large dissipation and Joule heating. 

A new and alternative current-induced spin-torque mechanism has been observed in systems with broken spatial inversion symmetry and strong spin-orbit coupling (SOC).~\cite{Chernyshov:nature09, Miron:nature10, Fang:nn2013, Fan:nc2013,Kurebayashi:arxiv2013, Chiara:nn2015, SOT:Review} Due to the SOC of these systems, an electric current is always accompanied by a net spin-polarization of the charge carriers,~\cite{ Bernevig:prb05,  Manchon:prb08,  Garate:prb09,  Hals:epl10, Pesin:prb2012, Bijl:prb2012, Wang:prl2012, Freimuth:prb2014,   Hals:prb2013b} which via the exchange interaction produces a torque on the magnetization. Such relativistic current-induced torques are commonly referred to as spin-orbit torques (SOTs). In contrast to the STTs, the SOTs require neither spin-polarizers nor textured ferromagnets to induce magnetization dynamics. Furthermore, the SOTs show remarkably high torque efficiencies,~\cite{Chernyshov:nature09, Miron:nature10, Fang:nn2013} leading to magnetization reversal at current densities that are approximately an order-of-magnitude smaller than that typically observed for STT-induced switching in metallic systems. 

\begin{figure}[ht] 
\centering 
\includegraphics[scale=1.0]{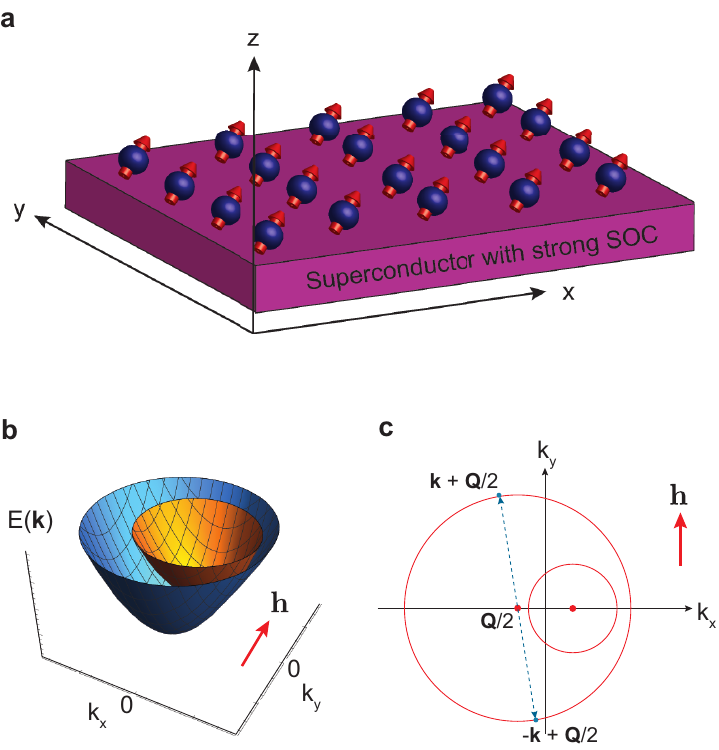}  
\caption{(color online). (a) Ferromagnetically ordered adatom spins on a superconductor with strong SOC.
(b) Energy dispersion of a system with Rashba SOC and an exchange field along the $y$-axis.
A typical Fermi surface of this system is illustrated in (c).  Optimal Cooper pairing occurs for momentum states with a finite center of mass momentum, i.e., $\mathbf{k} + \mathbf{Q}/2$ and $-\mathbf{k} + \mathbf{Q}/2$.   }
\label{Fig1} 
\end{figure} 

In the past few years, there has been a rapidly growing interest in superconductor-ferromagnet heterostructures with strong SOC.~\cite{Nadj-Perge:PRB2013,  Klinovaja:PRL2013, Vazifeh:PRL2013,  Simon:PRL2013, Nadj-Perge:Science2014, Sato:prb2010, Sau:prl2010, Mao:prb2010, Bjornson:prb2013,  Li:arxiv2015} These systems are particularly intriguing because they are considered to be promising platforms for realizing topological superconductivity. This was recently experimentally demonstrated for a chain of magnetic atoms placed on a conventional superconductor, where signatures of Majorana fermions at the edges of the chain were observed.~\cite{Nadj-Perge:Science2014} Topological superconductivity has also been predicted in two-dimensional lattices of ferromagnetically ordered adatoms.~\cite{Sato:prb2010, Sau:prl2010, Mao:prb2010, Bjornson:prb2013,  Li:arxiv2015} 

So far, there is little knowledge of how supercurrents in these systems influence the ordered spins. However, studies on ferromagnetic Josephson junctions have shown that supercurrents can produce a torque on the ferromagnetic interlayer via the creation of spin-triplet Cooper pairs.~\cite{Waintal:prb2002, Zhao:prb2002, Konschelle:prl2009, Linder:prl2012, Linder:prb2014, Halterman:prb2015, Linder:review} The spin-triplet correlations in the interlayer are generated either via magnetic textures, ferromagnet-normal metal-ferromagnet trilayers, or SOC.  Superconductors with broken spatial inversion symmetry and strong SOC will be in a mixed superconducting state of singlet and triplet pairings.~\cite{Gorkov:prl2001} The Cooper pairs can thus develop a net spin polarization when the time reversal symmetry is broken. Because the superconductor/adatom systems have both broken spatial inversion symmetry and strong SOC, an interesting question is whether supercurrents in these systems result in current-driven magnetization dynamics.
 
In this work, we consider a lattice of ferromagnetically ordered spins in contact with a conventional superconductor with broken spatial inversion symmetry and strong SOC (Fig.~\ref{Fig1}a). We find that supercurrents produce a reactive SOT on the spins  and formulate a phenomenological description of the supercurrent-induced magnetization dynamics. In contrast to normal metallic ferromagnets, the back-action of the spin dynamics on the superconducting system is important, and the resulting equations for the condensate and the spin system should be solved simultaneously to provide a correct description of the dynamics. Furthermore, we study the spin-torque mechanism  by using a tight-binding Bogoliubov-de Gennes (BdG) formalism to self-consistently calculate the response of the system to a supercurrent.  We show that the SOT originates from current-induced spin-triplet correlations, which is determined by the orientation of the supercurrent with respect to the crystallographic axes. Moreover, we find that there exists an intrinsic limitation for the maximum achievable SOT and estimate the corresponding effective SOT field to be on the order of $0.16$ mT in proximitized hole-doped semiconductors, which is comparable to the critical SOT field for magnetization reversal in (Ga,Mn)As. We therefore believe that the supercurrent-induced SOTs can lead to the development of new efficient techniques for manipulating magnetization, which minimize the disadvantages associated with dissipation and Joule heating.

\section{Phenomenological Description} 
In what follows, we develop a phenomenology that captures the low-frequency long-wavelength physics of the supercurrent-induced magnetization dynamics. The superconducting condensate is treated in the framework of the Ginzburg-Landau theory, which is valid at length scales larger than the superconducting coherence length $\zeta_0$. We assume a homogeneous ferromagnetic equilibrium state and consider spatial modulations of the ferromagnetic order parameter at length scales much larger than the exchange length $l_{ex}\sim \sqrt{J/K}$ set by the spin stiffness $J$ and relevant anisotropy constants $K$. Typically, $l_{ex}\sim 10 - 100$ nm and $\zeta_0\sim 40-360$ nm.\cite{Brataas:nm2012, deGennes:book} The characteristic frequency $\omega$ of ferromagnets is on the order of $\omega\sim 1$ GHz,~\cite{Brataas:nm2012}  which is far below the typical energy gap of superconductors: $\hbar\omega << \Delta \sim 0.18-1.5$ meV.\cite{deGennes:book} The magnetization precession will therefore not lead to quasi-particle excitations in the superconductor and we can assume that the condensate responds adiabatically to the magnetization dynamics. 

We start by formulating the free energy functional, $F \left[ \mathbf{m}, \psi, \mathbf{A} \right]$, of the system:
\begin{equation}
F= \int {\rm d}\mathbf{r}\left[ \mathcal{F}_m \left( \mathbf{m} \right) + \mathcal{F}_{me} \left( \mathbf{m},\psi, \mathbf{A} \right) + \mathcal{F}_e \left( \psi, \mathbf{A} \right)  \right].  \label{Eq:Ftot}
\end{equation}
Here, $\mathbf{m} (\mathbf{r}, t)$ represents the order parameter of the spin system and is a unit vector parallel to the magnetization $\mathbf{M} (\mathbf{r}, t)= M_s \mathbf{m}(\mathbf{r}, t)$, $\psi (\mathbf{r}, t)$ is the order parameter field of the superconductor, and $\mathbf{A} (\mathbf{r}, t)$ is the magnetic vector potential that yields the magnetic field $\mathbf{B}(\mathbf{r}, t)= \boldsymbol{\nabla}\times \mathbf{A} (\mathbf{r}, t)$.
$\mathcal{F}_{m}$ and $\mathcal{F}_{e}$ are the free energy densities of the isolated spin system and superconducting condensate, respectively,~\cite{Brataas:nm2012,deGennes:book}
\begin{eqnarray}
\mathcal{F}_{m} &=& \sum_{ij} \frac{J_{ij}}{2}\partial_i \mathbf{m}\cdot \partial_j \mathbf{m} + U(\mathbf{m}),  \nonumber \\
\mathcal{F}_{e} &=& \sum_{ij} K_{ij} (\Pi_i\psi)^{\ast}( \Pi_j\psi) + \alpha |\psi|^2 + \frac{\beta}{2} |\psi|^4 + \frac{B^2}{ 8\pi} , \nonumber
\end{eqnarray}
where $U$ describes the magnetic anisotropy energy,  $\boldsymbol{\Pi} = -i \hbar \boldsymbol{\nabla} - (2 e/c) \mathbf{A}$ is the momentum operator of the condensate, $2e$ is the charge of the Cooper pairs, and $c$ is the speed of light. 
$J_{ij}$ and $K_{ij}$ are second-rank polar tensors, which are invariant under the symmetry point group of the system.~\cite{comment:tensor}

The term $\mathcal{F}_{me}$ describes the coupling between the superconductor and the magnetization. 
We consider weak modulations of a homogenous ferromagnetic equilibrium state and can thus neglect magnetoelectric coupling effects associated with magnetic textures. 
In this case, $\mathcal{F}_{me}$ is governed 
by the Lifshitz invariant~\cite{Edelstein:96}
\begin{equation}
\mathcal{F}_{me}= -\sum_{ij} \kappa_{ij} m_i  \Lambda_j ,  \label{Eq:Fme}
\end{equation}
where $\boldsymbol{\Lambda}=  \psi^{\ast} \boldsymbol{\Pi} \psi + \psi \boldsymbol{\Pi}^{\ast} \psi^{\ast}$ represents the momentum density of the superconducting condensate. 
The tensor $\kappa_{ij}$ is linear in the SOC and is an invariant axial tensor of the point group.~\cite{comment:tensor}
Consequently, the tensor vanishes for systems with spatial inversion symmetry. 
$\mathcal{F}_{me}$ can be derived microscopically by considering an s-wave superconductor with SOC of the form $\eta_{{\rm so}, ij}\sigma_i p_j $ (where $\eta_{{\rm so}, ij}\propto \kappa_{ij}$ ) and calculate the energy change due to a Zeeman field.~\cite{Edelstein:96}

So far, most works have concentrated on the effects of the Lifshitz invariant \eqref{Eq:Fme} in non-centrosymmetric superconductors exposed to an external magnetic field. 
However, two recent studies showed that $\mathcal{F}_{me}$ leads to persistent currents in a conventional superconductor with SOC when magnetic impurities are placed at the surface.~\cite{Balatsky:prl15}  

$\mathcal{F}_{me}$ couples the momentum of the condensate to the direction of the magnetization and favors a spatial modulation of $\psi$ in equilibrium.
The physical origin of $\mathcal{F}_{me}$ is an SOC-induced shift of the Fermi surface, leading to a finite center of mass momentum of the Cooper pairs. 
To illustrate this phenomenon, consider a system with Rashba SOC and a Zeeman splitting $h_0$ along the $y$-axis  induced by the magnetization (Fig.~\ref{Fig1}b): $H(\mathbf{k})= \hbar^2k^2 /2m + \alpha_R (\mathbf{k}\times\hat{\mathbf{z}} )\cdot \boldsymbol{\sigma} + h_0\sigma_y$. Here, $\boldsymbol{\sigma}$ is a vector consisting of the Pauli matrices, $m$ is the effective quasi-particle mass, and $\alpha_R$ parameterizes the SOC.  
 For this system, the Lifshitz invariant becomes $\mathcal{F}_{me}= -\kappa (\hat{\mathbf{z}}\times\mathbf{m})\cdot  \boldsymbol{\Lambda}$.
The Fermi surface of the Hamiltonian is two circles, whose centers are shifted in opposite  directions along the x-axis (Fig.~\ref{Fig1}c). Due to the shift of the Fermi surface, the optimal Cooper pairing occurs for momentum states with a finite center of mass momentum, i.e., $\mathbf{k} + \mathbf{Q}/2$ and $-\mathbf{k} + \mathbf{Q}/2$. Therefore, the order-parameter field gains a spatial modulation $\psi\sim \exp(i\mathbf{Q}\cdot \mathbf{r})$ in equilibrium; a state
that is referred to as the helical phase.~\cite{Edelstein:96} Phenomenologically, the helical phase is captured by the Lifshitz invariant  $\mathcal{F}_{me}\sim -\kappa (\hat{\mathbf{z}}\times\mathbf{m})\cdot  \mathbf{Q}$, which favors the vector $\mathbf{Q}$ to be perpendicular to the in-plane component of the magnetization. 

In what follows, we demonstrate that the Lifshitz invariant also leads to a reciprocal phenomenon of the helical phase. If $\psi$ is forced to have a spatial modulation $\sim \exp(i\mathbf{q}\cdot \mathbf{r})$  such that a supercurrent is induced, then the condensate can via  $\mathcal{F}_{me}$ lower its energy by developing a net spin density $\mathbf{S}_{\rm ind}$ (and magnetic moment $\mathbf{m}_{\rm ind}$) perpendicular to the vector $\mathbf{q}$: $\mathcal{F}_{me}\sim -\kappa (\hat{\mathbf{z}}\times\mathbf{m}_{\rm ind} )\cdot  \mathbf{q} < 0$. Importantly, we find that the induced spin density $\mathbf{S}_{\rm ind}$ produces a novel SOT on the magnetization.    

The magnetization dynamics is described by the Landau-Lifshitz-Gilbert (LLG) equation~\cite{Brataas:nm2012}
\begin{equation}
\dot{\mathbf{m}} = -\gamma \mathbf{m}  \times \left[ \mathbf{H}_{\rm eff} + \mathbf{H}_{\rm so} \right] + \alpha_G \mathbf{m}  \times  \dot{ \mathbf{m}}  . \label{Eq:LLG}
\end{equation}
Here, $\mathbf{H}_{\rm eff}= -(1/M_s) \delta F_m / \delta \mathbf{m}$ is the effective field found from the magnetic free energy functional $F_m= \int {\rm d}\mathbf{r} \mathcal{F}_m$, $\gamma$ is the gyromagnetic ratio, and the term proportional to the Gilbert damping parameter $\alpha_G$ determines the magnetization dissipation.  Because of the Lifshitz invariant~\eqref{Eq:Fme}, the variation of Eq.~\eqref{Eq:Ftot} with respect to the magnetization also yields a reactive SOT-field 
\begin{equation}
H_{\rm so, i} = \sum_j \kappa_{ij}\Lambda_j / M_s,  \label{Eq:Hso}
\end{equation}
which is governed by the SOC and the momentum density of the superconducting condensate.

The magnetization evolves slowly on the characteristic timescale of the electron dynamics. We can therefore assume that the superconducting condensate at time $t$ is close to the equilibrium state with the 
static magnetization $\mathbf{m} (t)$. The equilibrium state, which is determined by 
the Ginzburg-Landau (GL) equations, is obtained by variational minimization of the free energy \eqref{Eq:Ftot}. 
The variation with respect to $\psi^{\ast}$ yields the equation 
\begin{equation}
\sum_{ij} K_{ij} \Pi_i \Pi_j \psi +  \alpha \psi  + \beta |\psi |^2 \psi - 2 \kappa_{ij}m_i \Pi_j\psi = 0 ,
\label{Eq:GL1}
\end{equation}
whereas  a variation of $\mathbf{A}$ provides the equation 
\begin{eqnarray}
\frac{j_{s,i}}{2e} &=& \sum_j K_{ij}\psi^{\ast} \Pi_j\psi +  K_{ji}\psi \Pi_j^{\ast} \psi^{\ast}   -  2  \kappa_{ji}m_j |\psi |^2 . 
\label{Eq:GL2}
\end{eqnarray}
Here, $\mathbf{j}_s= (c/4\pi) \left(\boldsymbol{\nabla}\times \mathbf{B} \right)$ is the supercurrent density.
The conventional GL equations are obtained for a fully isotropic system, in which $K_{ij}= K \delta_{ij}$ and $\kappa_{ij}=0$. 

Eqs.~\eqref{Eq:LLG}, \eqref{Eq:GL1} and \eqref{Eq:GL2} give a phenomenological description of the coupled dynamics of the spin system and the superconducting condensate.   
Via the Lifshitz invariant, the state of the superconducting condensate depends on the direction of the magnetization. 
The effects of $\mathcal{F}_{me}$ become crucially important when the length scale $2\pi / Q$ associated with the helical wavevector $Q$ is
smaller than the characteristic length scales of the ferromagnetic system.
In this case, the condensate is strongly affected by the magnetization dynamics and its state cannot be considered as quasi-static for the dynamics.
Thus, Eqs.~\eqref{Eq:LLG}, \eqref{Eq:GL1} and \eqref{Eq:GL2}  should be solved simultaneously to provide a correct description of both the magnetization dynamics and the superconducting condensate.
This differs markedly from the situation in normal metallic ferromagnets, in which the back-action of the spin dynamics on the itinerant electron system usually can be disregarded in the 
solution of the LLG equation. 

For a two-band model with Rashba SOC, the helical wavevector is on the order of  $Q\sim \delta N h_0/\hbar v_F$ where $v_F$ is the Fermi velocity.\cite{HelicalQ} Here, the factor $\delta N= (N_{+} - N_{-})/ (N_{+} - N_{-})$ measures of the difference between the density of states $N_{\pm}$ of the two bands at the Fermi energy. In the limit $\alpha_R/v_F << 1$, it is determined by $\delta N = 2 \alpha_R / \hbar v_F$.
To get some insight into the typical scale of $Q$, let us estimate $Q$ for the proximitized hole-doped semiconductor system studied in Sec.~\ref{Sec:Micro}.
For this system, we find the helical wavevector $Q\sim 1.7\times 10^{7}$ m$^{-1}$ (material parameters are given in Sec.~\ref{Sec:Model}). 
This is about an order of magnitude larger than the wavevector observed for the pair potential in a proximitized HgTe quantum well system subjected to an in-plane magnetic field of $1$ T.~\cite{Hart:arxiv2015}
Thus, it is likely that the spatial modulation of the order parameter field $\psi$ becomes important for the magnetization dynamics at length scales larger than $0.1-1.0$~$\mu$m.       

\section{Microscopic Calculation} \label{Sec:Micro}
To gain a better understanding of the underlying physical mechanisms of the SOT, we will now use the BdG formalism to self-consistently calculate the response of the system to a supercurrent. 

\subsection{Model} \label{Sec:Model}
We model the two-dimensional superconductor by the tight-binding Hamiltonian
\begin{eqnarray}
H &=& -\tilde{t} \sum_{\langle \mathbf{ij} \rangle} \mathbf{c}_{\mathbf{i}}^{\dagger} \mathbf{c}_{\mathbf{j}} - \mu\sum_{\mathbf{i}} \mathbf{c}_{\mathbf{i}}^{\dagger} \mathbf{c}_{\mathbf{i}} + \sum_{\mathbf{i}}  \mathbf{c}_{\mathbf{i}}^{\dagger} \left(  \mathbf{h}_{\mathbf{i}} \cdot\boldsymbol{\sigma} \right) \mathbf{c}_{\mathbf{i}} + \label{Eq:H0} \\
& &   i \sum_{\langle \mathbf{ij} \rangle}  \mathbf{c}_{\mathbf{i}}^{\dagger}  \left(  \boldsymbol{\sigma} \cdot \boldsymbol{\eta}_{\rm so}\cdot \hat{\mathbf{d}}_{\mathbf{ij}} \right) \mathbf{c}_{\mathbf{j}} + \sum_{\mathbf{i}} \left( \Delta_{\mathbf{i}}  c_{\mathbf{i}\uparrow}^{\dagger} c_{\mathbf{i}\downarrow}^{\dagger} +  h.c. \right)  .    \nonumber
\end{eqnarray}
Here, $\mathbf{c}^{\dagger}_{\mathbf{i}}  = (c_{\mathbf{i} \uparrow}^{\dagger}  \  c_{\mathbf{i} \downarrow}^{\dagger} )$, where $c_{\mathbf{i} \tau}^{\dagger} $ is a fermionic creation operator that creates a particle with spin $\tau$ at lattice site $\mathbf{i}= (x,y)$ and the symbol $\langle \mathbf{ij} \rangle$ implies a summation over nearest lattice sites.  $\tilde{t}$ is the spin-independent hopping energy,  $\mathbf{h}_{\mathbf{i}}$ is the Zeeman splitting induced by the adatom spins ($\mathbf{h}_{\mathbf{i}}$ and $ \mathbf{m} (\mathbf{i})$ are collinear), and $ \hat{\mathbf{d}}_{\mathbf{ij}} $ is a unit vector that points from site $\mathbf{j}$ to site $\mathbf{i}$. 

The second-rank tensor $(\eta_{\rm so})_{ij}$ parameterizes the SOC. We consider a system described by the $C_{2v}$ point group, in which the SOC can be decomposed in two terms having Rashba and Dresselhaus symmetry, respectively. For Rashba SOC, the SOC tensor takes the form $\boldsymbol{\eta}_{\rm so}= \tilde{\alpha}_{R}i\sigma_y$, whereas the specific form the Dresselhaus SOC depends on how the coordinate system is fixed with respect to the crystallographic axes.
If the $x$-axis is along one of the two reflection planes of $C_{2v}$, then the Dresselhaus SOC tensor is  $\boldsymbol{\eta}_{\rm so}= \tilde{\alpha}_{D}\sigma_x$. 
With respect to this reference frame, a rotation of the axes by $\pi/2$ degrees about  the $z$-axis leads to a sign change of $\tilde{\alpha}_D$, while a rotation of $\pi/4$ degrees changes the tensor to $\boldsymbol{\eta}_{\rm so}= \tilde{\alpha}_{D}\sigma_z$.  Note that $(\eta_{\rm so})_{ij}$ and $\kappa_{ij}$ satisfy the same transformation rules and thus have the same tensorial forms, i.e., $(\eta_{\rm so})_{ij}\propto \kappa_{ij} $.

\begin{figure}[ht] 
\centering 
\includegraphics[scale=1.0]{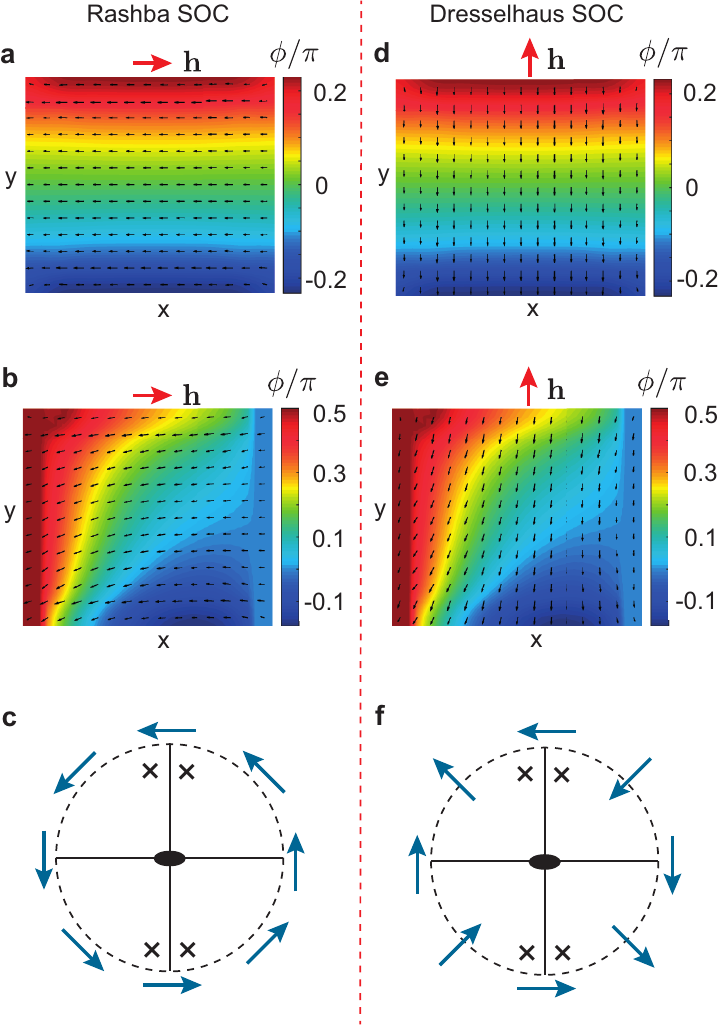}  
\caption{(color online). (a) The equilibrium state for a system with Rashba SOC and a Zeeman splitting field along $x$. 
The color represents the phase $\phi_{\mathbf{i}}$ of the pair potential $\Delta_{\mathbf{i}} = |\Delta_{\mathbf{i}}|\exp (i \phi_{\mathbf{i}} )$, while
the black arrows illustrate the local spin density  $\mathbf{S} (\mathbf{i})= (\hbar/2) \langle \mathbf{c}^{\dagger}_{\mathbf{i}} \boldsymbol{\sigma} \mathbf{c}_{\mathbf{i}} \rangle$.
(b) System (a) with an enforced superconducting phase difference of $\pi/2$ between two of the sample edges. 
(c) Symmetry plot of the induced spin density for a Rashba system. The figure shows the stereographic projection of the $C_{2v}$ point group and the blue arrows illustrate the orientation of the induced 
spin density for a supercurrent along different crystallographic directions.     
(d)-(f) Show corresponding plots for the case with Dresselhaus SOC and a Zeeman splitting field along $y$.
In (a)-(b) and (d)-(e), the size of the system is $31\times 27$ grid points. }
\label{Fig2} 
\end{figure} 

$\Delta_{\mathbf{i}}= V \langle c_{\mathbf{i} \uparrow} c_{\mathbf{i} \downarrow}  \rangle$ describes the superconducting s-wave pairing and is determined by
\begin{eqnarray}
\Delta_{\mathbf{i}} &=& -\frac{V}{2}\sum_{n\tau\tau^{'}} (i\sigma_y)_{\tau \tau^{'}} v_{n\tau}^{\ast} (\mathbf{i}) u_{n\tau^{'}} (\mathbf{i}) \left[ 1-2f (\epsilon_n) \right] .\label{Eq:delta1}
\end{eqnarray}
Here, $V>0$ is the on-site attractive interaction between the quasi-particles, $\langle ... \rangle$ denotes the thermal average, $f(\epsilon)$ is the Fermi-Dirac distribution,
and we have inserted the Bogoliubov transformation  $c_{\mathbf{i}\tau}= \sum_n [  u_{n\tau} (\mathbf{i}) \gamma_n +  v_{n\tau}^{\ast} (\mathbf{i}) \gamma_n^{\dagger} ]$, where
 $\gamma_n^{\dagger}$ ($\gamma_n$) are the Bogoliubov quasi-particle creation (destruction) operators, which represent a complete set of energy eigenstates:  $H= E_g + \sum_n \epsilon_n \gamma_n^{\dagger}\gamma_n $.
$E_g$ is the groundstate energy; the summation runs over positive energy eigenstates with an energy smaller than the cut-off energy $\hbar\omega_D$  set by the Debye frequency $\omega_D$.

The Hamiltonian \eqref{Eq:H0} is transformed to BdG Hamiltonian by using the Bogoliubov transformation~\cite{deGennes:book}, which is then iteratively solved together with the self-consistency condition \eqref{Eq:delta1}~\cite{Sacramento:prb07}  
until the Euclidean norm of the pair potential ($\| \Delta \| = \sqrt{\sum_{\mathbf{i}} |\Delta_{\mathbf{i}}  |^2}$) reaches a relative error on the order $10^{-5}$.  
In the following, the Hamiltonian \eqref{Eq:H0} is scaled with the hopping energy $\tilde{t}$ and the chemical potential, the pairing strength, the Rashba (Dresselhaus) SOC, the Zeeman splitting, the thermal energy $k_B T$, and the Debye frequency are set to: $\mu/ \tilde{t}= -4$, $V/ \tilde{t} = 5$,  $\tilde{\alpha}_{R(D)}/ \tilde{t} = 0.5$, $h_0/ \tilde{t}= 0.1$, $k_B T/ \tilde{t}= 0.001$, and $\hbar \omega_D / \tilde{t}= 2.0$, respectively. In Eq.~\eqref{Eq:H0}, we use open boundary conditions. 
The hopping and Rashba energies in the tight-binding Hamiltonian \eqref{Eq:H0} are related to a central difference discretization of the corresponding continuum model via the relationships 
$ \tilde{t}= \hbar^2/2ma^2$ and $\tilde{\alpha}_{R (D)}/ \tilde{t} = ma\alpha_{R(D)}/\hbar^2$, where $a$ is the spacing between the grid points and $\alpha_{R(D)}$ is the SOC parameter in the continuum model.
The parameter values given above model a lightly hole-doped semiconductor in proximity to a conventional s-wave superconductor, in which 
the effective mass is $m=0.6 m_e$ ($m_e$ is the electron mass), the SOC is $\alpha_{R(D)}= 0.21$~eV\AA, the Fermi energy is $E_F= 2.47$~meV when measured from the bottom of the lowest subband, and 
the Fermi wavelength is $\lambda_F \sim 20$ nm, which is much larger than the discretization constant $a= 3$~nm.~\cite{Stormer:prl83}

\begin{figure}[tp] 
\centering 
\includegraphics[scale=1.0]{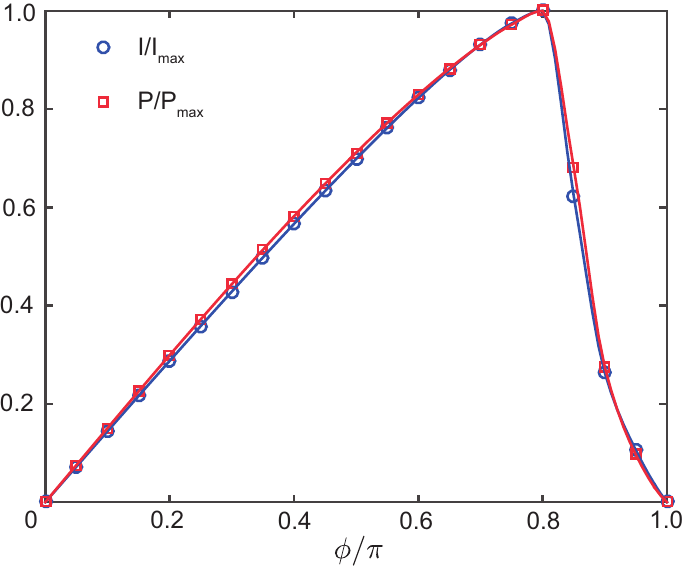}  
\caption{(color online). Current-phase relation and the induced spin-polarization of the Cooper pairs for the Rashba system in Figs.~\ref{Fig2}a,b. The lines represent piecewise polynomial fits of the data points.}
\label{Fig3} 
\end{figure} 

\subsection{Results and discussion}
First, we study the equilibrium spin density $\mathbf{S} (\mathbf{i})= (\hbar/2) \langle \mathbf{c}^{\dagger}_{\mathbf{i}} \boldsymbol{\sigma} \mathbf{c}_{\mathbf{i}} \rangle$ of the superconducting condensate.
We consider the two cases with Rashba and Dresselhaus SOC separately. Fig.~\ref{Fig2}a  shows the self-consistent solution for a Rashba system with an exchange field along $x$. The black arrows represent the spin density, while the color illustrates the phase $\phi_{\mathbf{i}}$ of the pair potential  $\Delta_{\mathbf{i}} = |\Delta_{\mathbf{i}}|\exp (i \phi_{\mathbf{i}} )$.  The phase variation perpendicular to the exchange field (i.e., along $y$) is a signature of the helical phase. We see that the condensate has a net spin polarization anti-parallel to the exchange field. This is also the case for a system with Dresselhaus SOC  of the form $\boldsymbol{\eta}_{\rm so}= \tilde{\alpha}_{D}\sigma_z$  and an exchange field along $y$ (Fig.~\ref{Fig2}d). Note that in this case, the pair potential has a phase variation parallel to the exchange field, which is in agreement with Eq.~\eqref{Eq:Fme} when $\kappa_{ij}\propto (\sigma_z)_{ij}$.

Next, we investigate the effects of a supercurrent. A supercurrent is induced along the $x$-axis by enforcing the pair potential to have a constant phase in a small region close to each of the two boundaries along $x$. We set the widths of these two regions to three lattice points. Thus, the pair potential is solved self-consistently for the entire sample except for the two regions at the boundaries where the phase $\phi_{\mathbf{i}}$ is kept fixed (however, the magnitude $|\Delta_{\mathbf{i}}|$ is allowed to optimize itself). These two regions will therefore act as sinks/sources for the supercurrent. 

Fig.~\ref{Fig2}b,e shows the solution for the Rashba and Dresselhaus systems with a phase difference of $\pi/2$ between the two boundaries. In both cases,  the spin density is tilted away from the equilibrium value. In other words: the supercurrent induces  a spin-density $\mathbf{S}_{\rm ind}$. A similar inverse spin-galvanic effect has been theoretically predicted for superconductors with Rashba SOC in the absence of magnetization.~\cite{Edelstein:prl95, Edelstein:prb05}

$\mathbf{S}_{\rm ind}$ is solely an effect of the SOC, and its orientation is determined by the direction of the supercurrent relative to the crystallographic axes. 
Fig.~\ref{Fig2}c,f shows the stereographic projection of the $C_{2v}$ point group, and the blue arrows illustrate the orientation of $\mathbf{S}_{\rm ind}$ for different directions of the supercurrent ($\nabla \phi > 0$ along the different directions). 
Generally, the supercurrent results in a spin density $S_{\rm ind, i} \propto \kappa_{ij} \Lambda_j$. 
Via the exchange coupling,  $\mathbf{S}_{\rm ind}$ produces a torque on the magnetization and is the physical origin of the SOT field in Eq.~\eqref{Eq:Hso}: $\mathbf{H}_{\rm so} \propto \mathbf{S}_{\rm ind}$.   

The polarization of the condensate originates from spin-triplet correlations.  
Let  $g_{T_{+}} (\mathbf{i})=  \langle \tilde{c}_{\mathbf{i} \uparrow} \tilde{c}_{\mathbf{i} \uparrow} \rangle$  ($g_{T_{-}} (\mathbf{i})=  \langle \tilde{c}_{\mathbf{i} \downarrow} \tilde{c}_{\mathbf{i} \downarrow} \rangle$) denote the amplitude for triplet pair correlations with spin up (down) along an arbitrary quantization axis, which is determined by the unitary rotation operator $U_{\tau \tau^{'}}$. Here,  $\tilde{c}_{\mathbf{i} \tau}= U_{\tau \tau^{'}} c_{\mathbf{i} \tau^{'}}$ are the fermionic operators in the rotated frame.
The quantity  $P= \sum_{\mathbf{i}} [ |g_{T_{+}} (\mathbf{i})|^2 - |g_{T_{-}} (\mathbf{i})|^2  ]$ represents a measure of the spin polarization of the Cooper pairs along the quantization axis.
In Fig.~\ref{Fig3}, we consider the Rashba system in Fig.~\ref{Fig2}a-b and plot $P$ and the supercurrent $I$ along $x$ as a function of the phase difference $\phi$ between the left and right boundaries.
The spin quantization axis is along $y$. 
It is clear from Fig.~\ref{Fig3} that $P$ is proportional to the supercurrent. 
We obtain a similar relationship between the current and $P$ for the Dresselhaus system in Fig.~\ref{Fig2}d-e when the polarization is measured along $x$.  
Thus, we conclude that the underlying physical mechanism of the SOT field \eqref{Eq:Hso} is current-induced spin-triplet correlations.

\begin{figure}[tp] 
\centering 
\includegraphics[scale=1.0]{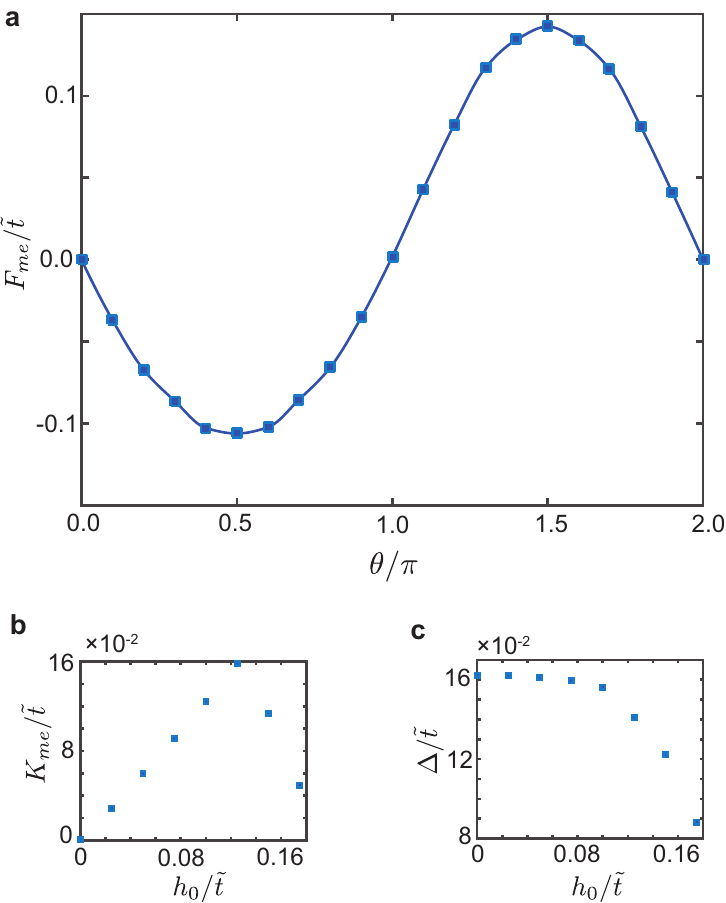}  
\caption{(color online). (a) A microscopic calculation of the anisotropic part $F_{me} (\theta)= F_s (\theta) - F_s (0)$ of the superconductor's free energy $F_s$ for different directions of  $\mathbf{h} = h_0 [\cos (\theta), \sin (\theta), 0]$.
(b) The  magnetoelectric anisotropy constant $K_{me}= (|F_{me} (\pi / 2)| + |F_{me} (3\pi / 2)|)/2$ for different values of $h_0$.
(c) The average energy gap for different values of $h_0$.
In all figures, the size of the system is $25\times 23$ grid points and the phase difference between the two boundaries is $\phi=0.8\pi$. 
The squares represent the calculated values, while the line in (a) is a piecewise polynomial fit.
In (a) the free energy was calculated for $h_0/\tilde{t}=0.1$.  }
\label{Fig4} 
\end{figure} 

The strength of the SOT field can be investigated by self-consistently calculating the free energy of the condensate for different directions of $\mathbf{h} = h_0 [\cos (\theta), \sin (\theta), 0]$. Here, $\theta$ is the angle with the $x$-axis, which is parallel to the direction of the supercurrent. The anisotropic part of the free energy is then a direct measure of the Lifshitz invariant  \eqref{Eq:Fme}.  

We consider a system with Rashba SOC. The free energy of an inhomogeneous superconductor is~\cite{Kosztin:prb98}
\begin{equation}
F_s= -\frac{1}{\beta}\sum_n {\rm ln} \left[ 2 \cosh \left( \frac{ \beta \epsilon_n }{2} \right)  \right] + \frac{1}{V} \int {\rm d}\mathbf{r} |\Delta ( \mathbf{r} )|^2 ,  \label{Eq:Fs}
\end{equation}
where the sum is over the positive energy eigenstates and $\beta= 1/ k_B T$. 

Fig.~\ref{Fig4}a shows the anisotropic part $F_{me} (\theta)= F_s (\theta) - F_s (0)$ of the free energy. The angular dependence of $F_{me}$ follows the functional form $F_{me} \sim (\hat{\mathbf{z}} \times \mathbf{h})\cdot \boldsymbol{\Lambda}$,  which is consistent with the Lifshitz invariant \eqref{Eq:Fme} when a current is applied along the $x$-axis (with extrema at $\theta = \pi/2$ and $\theta= 3\pi /2$). The different extremum values  at $F_{me} (\pi / 2)$ and $F_{me} (3 \pi / 2)$ is caused by a change in the momentum density due to the helical modulation (along the $x$-axis) of the order parameter field. 

The effect of the Zeeman splitting $h_0$ on the SOT is twofold. Firstly, it determines the coupling strength between the spin system and the condensate and thus enhances the magnetoelectric coupling $F_{me}$. Secondly, it suppresses superconductivity and thus reduces the supercurrent/momentum density. The competition between these two counteracting effects implies that there exists an intrinsic limitation for the maximum achievable SOT. Fig.~\ref{Fig4}b shows the magnetoelectric anisotropy constant $K_{me}= (|F_{me} (\pi / 2)| + |F_{me} (3\pi / 2)|)/2$. A maximum SOT is achieved for $h_0 / \tilde{t} \sim 0.125$ with $ K_{me} \sim  0.16 \tilde{t}= 1.13$~meV and corresponds the point where the Zeeman splitting  is comparable to the pair potential, i.e., $h_0\sim \Delta$. For larger values of $h_0$, the suppression of the superconductivity becomes stronger (Fig.~\ref{Fig4}c), which leads to a lowering of $K_{me}$. 

The effective SOT field induced by the supercurrent is $H_{so}\sim K_{me}/VM_s$, where $V$ is the volume of the ferromagnetic system.
Assuming $M_s= 70.8$ e.m.u. cm$^{-3}$,~\cite{Chiara:nn2015}  $V= 23\times 25 a^3$, and $K_{me}= 1.13$~meV, yields an SOT field on the order of $H_{so}\sim 0.16$ mT. 
In the ferromagnetic semiconductor (Ga,Mn)As, current-driven magnetization switching has been observed for effective SOT fields on the order of $0.14-0.35$ mT.~\cite{Chernyshov:nature09} 
Therefore, it is reasonable to believe that the supercurrent-induced SOT is strong enough to manipulate the magnetization of the ferromagnetically ordered spins.

\section{Summary} 
In summary, we have studied the magnetization dynamics of a two-dimensional lattice of spins in contact with a conventional superconductor and have formulated a phenomenological description of the coupled dynamics of the superconducting condensate and the magnetization. Interestingly, we found that  supercurrents induce a reactive SOT field that originates from current-induced spin-triplet correlations and whose spatial orientation is determined by the symmetry of the SOC.
Furthermore, we showed that there exists an intrinsic limitation for the maximum achievable SOT, which is determined by the coupling strength between the condensate and the spin system. 
Based on material parameters for a proximitized hole-doped semiconductor, we estimated the induced SOT field to be on the order of $0.16$ mT.

\appendix

\section{Expressions for spin-density, pair correlations and current density}\label{App1}
The Hamiltonian \eqref{Eq:H0} can be diagonalized by using the Bogoliubov transformation~\cite{deGennes:book} 
\begin{equation}
c_{\mathbf{i}\tau} (\mathbf{r} ) = \sum_n \left( u_{n\tau}(\mathbf{i} ) \gamma_n + v_{n\tau}^{\ast}( \mathbf{i} ) \gamma^{\dagger}_n  \right) . \label{Eq:BT}
\end{equation}
Here, $\gamma^{\dagger}_n$ and $\gamma_n$ are the Bogoliubov quasi-particle creation and destruction operators, which satisfy fermionic 
anti-commutation relations and represent a complete set of energy eigenstates:  
\begin{equation}
H= E_g + \sum_n \epsilon_n \gamma_n^{\dagger}\gamma_n .
\end{equation}
$E_g$ is the ground state energy, and the summation runs over positive energy eigenstates with an energy lower than the cut-off energy set by the Debye frequency.
The thermal averages of the Bogoliubov quasi-particle excitations are given by $ \langle \gamma_n^{\dagger} \gamma_n  \rangle = f(\epsilon_n)$,  
where $f(\epsilon)= 1/({\exp(\beta\epsilon)} + 1)$  is the  Fermi-Dirac distribution. It is also useful to introduce the distribution function of the corresponding hole states: $f_h(\epsilon) = 1- f(\epsilon)$. 

By using the Bogoliubov transformation \eqref{Eq:BT}, the spin density $\mathbf{S} (\mathbf{i} )= (\hbar/2) \langle \mathbf{c}^{\dagger}_{\mathbf{i}} \boldsymbol{\sigma} \mathbf{c}_{\mathbf{i}} \rangle$  can be expressed as
\begin{eqnarray}
S_{\alpha} (\mathbf{i}) &=& \frac{\hbar}{2} \sum_{\tau \tau^{'}} (\sigma_{\alpha})_{\tau \tau^{'}} \rho_{\tau \tau^{'}} (\mathbf{i}) , \nonumber  \\
\rho_{\tau \tau^{'}} (\mathbf{i})  &\equiv & \sum_n [ (u_{n \tau}^{\ast} (\mathbf{i})  u_{n \tau^{'}} (\mathbf{i})   - v_{n\tau} (\mathbf{i}) v_{n\tau^{'}}^{\ast} (\mathbf{i})     )f(\epsilon_n) \nonumber \\
& &  +  v_{n\tau} (\mathbf{i}) v_{n\tau^{'}}^{\ast} (\mathbf{i}) ]  . \nonumber
\end{eqnarray}

The charge density  $\rho_{\mathbf{i}}=  q \langle n_{\mathbf{i}}  \rangle $ at site $\mathbf{i}$ is given by the thermal average of the number operator $n_{\mathbf{i}}=  \mathbf{c}_{\mathbf{i}}^{\dagger} \mathbf{c}_{\mathbf{i}} $, where
$q$ is the charge of the quasi-particles.
An expression for the current density $\mathbf{j}_s (\mathbf{i})$ is found from the Heisenberg equation $dn_{\mathbf{i}}/dt= (i/\hbar) [H,n_{\mathbf{i}}]$, which yields 
\begin{widetext}
\begin{eqnarray}
\left( \mathbf{j}_s (\mathbf{i}) )\right)_k &=& \frac{2q \tilde{t} }{\hbar} \sum_{n\tau}  {\rm Im}\left[ u_{n \tau}^{\ast} (\mathbf{i}) D_k u_{n \tau} (\mathbf{i}) f (\epsilon_n)  + v_{n \tau} (\mathbf{i}) D_k v_{n \tau}^{\ast} (\mathbf{i}) f_h (\epsilon_n)   \right]  + \nonumber \\ 
& & \frac{2q}{\hbar} \sum_{n\tau\tau^{'} }  \left[ u_{n \tau}^{\ast} (\mathbf{i}) A_{k, \tau\tau^{'}} u_{n \tau^{'}} (\mathbf{i}) f (\epsilon_n)  + v_{n \tau} (\mathbf{i}) A_{k, \tau\tau^{'}}  v_{n \tau^{'}}^{\ast} (\mathbf{i}) f_h (\epsilon_n)   \right]  +
\frac{2q}{\hbar}  \sum_{\tau\tau^{'} }   {\rm Im} \left[ \Delta_{\mathbf{i}} i\sigma_{y, \tau\tau^{'}}  \langle c_{\mathbf{i}\tau }^{\dagger} c_{\mathbf{i}\tau^{'}}^{\dagger}  \rangle  \right]. \label{Eq:current}
\end{eqnarray}
\end{widetext}
Here, $D_k u_{n \tau} (\mathbf{i}) = [ u_{n \tau} (\mathbf{i} + \mathbf{a}_k) - u_{n \tau} (\mathbf{i} - \mathbf{a}_k) ] /2$ and 
$A_{k, \tau\tau^{'}}= \left(  \boldsymbol{\sigma} \cdot \boldsymbol{\eta}_{\rm so}\cdot \hat{\mathbf{d}}_{\mathbf{i} (\mathbf{i} - \mathbf{a}_k) } \right)_{\tau\tau^{'}}$, where $\mathbf{a}_k$ is the lattice vector along $k\in \left\{  x,y,z \right\}$. Note that the last term vanishes when the pair potential satisfies the self-consistency condition.  Otherwise, the term acts as a sink/source. Eq.~\eqref{Eq:current} is used to calculate the current-phase relation, which is shown in Fig. 3 of this article. 

The pair correlations are given by
\begin{eqnarray}
\langle c_{\mathbf{i}\tau} c_{\mathbf{i}\tau^{'}}  \rangle &=&  
\sum_n [ (v_{n \tau}^{\ast} (\mathbf{i})  u_{n \tau^{'}} (\mathbf{i})   - u_{n\tau} (\mathbf{i}) v_{n\tau^{'}}^{\ast} (\mathbf{i})  )f(\epsilon_n)  \nonumber \\
& & +  u_{n\tau} (\mathbf{i}) v_{n\tau^{'}}^{\ast} (\mathbf{i}) ] .
\end{eqnarray}

These correlation functions can be expressed in an arbitrary reference frame by transforming the fermionic operators: $\tilde{c}_{\mathbf{i} \tau}= U_{\tau \tau^{'}} c_{\mathbf{i} \tau^{'}}$.
$U_{\tau \tau^{'}} $ is the unitary rotation operator, which maps the $z$-axis to the quantization axis in the new reference frame.



\begin{thebibliography}{99}
\bibitem{comment:tensor}
For a symmetry transformation $\acute{\mathbf{r}}=\boldsymbol{\mathcal{R}}\mathbf{r}$,  the tensorial form of an invariant polar tensor $T_{ij\ldots k}$ of rank $k$ is determined by the equations
$ T_{ij\ldots k}= \mathcal{R}_{i\alpha} \mathcal{R}_{j\beta}\ldots \mathcal{R}_{k\gamma}T_{\alpha\beta\ldots \gamma}$,
whereas an invariant axial tensor is determined by
$T_{ij\ldots k}= |\mathcal{R}| \mathcal{R}_{i\alpha} \mathcal{R}_{j\beta}\ldots \mathcal{R}_{k\gamma}T_{\alpha\beta\ldots \gamma}$.
Here,  $|\mathcal{R}|$ represents the determinant of the matrix $\boldsymbol{\mathcal{R}}$, which is an element of the system's point group.
\end{thebibliography}
\end{document}